\documentclass[floatfix,prstab,showpacs,twocolumn]{revtex4}
\usepackage{graphicx}
\usepackage{dcolumn}
\usepackage{amsmath}
\usepackage{amsthm}

\begin{document}

\title{Benchmarking of 3D space charge codes using direct phase space
measurements from photoemission high voltage DC gun}

\author{Ivan V.~Bazarov} \email{ib38@cornell.edu}
\author{Bruce M.~Dunham}
\author{Colwyn Gulliford}
\author{Yulin Li}
\author{Xianghong Liu}
\author{Charles K.~Sinclair}
\author{Ken Soong}
\affiliation{
  Laboratory for Elementary Particle Physics,
  Cornell University, Ithaca, New York 14853
}
\author{Fay Hannon}
\affiliation{Lancaster University, Lancaster, United Kingdom}

\date{\today}

\begin{abstract}
We present a comparison between space charge calculations and direct
measurements of the transverse phase space for space charge dominated
electron bunches after a high voltage photoemission DC gun followed by
an emittance compensation solenoid magnet. The measurements were
performed using a double-slit setup for a set of parameters such as
charge per bunch and the solenoid current. The data is compared with
detailed simulations using 3D space charge codes GPT and Parmela3D
with initial particle distributions created from the measured
transverse and temporal laser profiles. Beam brightness as a function
of beam fraction is calculated for the measured phase space maps and
found to approach the theoretical maximum set by the thermal energy
and accelerating field at the photocathode. 
\end{abstract}

\pacs{29.25.Bx, 29.20.Ej, 29.27.Bd, 52.59.Sa, 52.65.Cc, 52.65.Rr, 52.70.Nc}
\keywords{photoinjectors, emittance, beam brightness, space charge
  dominated beams, space charge 3D simulations}
\maketitle

\section{Introduction\label{sec:intro}}

The generation of high-brightness electron beams remains the principle
challenge for a number of linear accelerator based projects including
the Energy Recovery Linac (ERL). Design of the electron sources relies
heavily on the use of space charge simulations \cite{Parmela, Astra,
  GPT, HOMDYN, Tredi, Trace3D}. Still, there remain a number of open
questions with regards to understanding and modeling of space charge
dominated bunched beams in photoemission guns. Different space charge
codes use varying degree of approximations to capture the most
significant physics relevant to beam dynamics in
photoinjectors. Certain codes allow self-consistent inclusion of
complex conducting boundaries at the expense of considerable increase
in the required computation time \cite{Mafia, Warp}, while many of
the mainstream codes widely used in the design of photoinjectors
assume open boundary conditions everywhere except in the vicinity of
the photocathode. Due to an inability to use direct self-force
calculations in a bunch with $\sim10^9$ particles, artificial smoothing of
the space charge potential is employed when tracking macroparticles,
which represent the actual bunch, either through meshing of the electron
cloud or via introduction of an effective size to the macroparticles
in a point-to-point calculation. As a result, and depending on the
simulation parameters, the space charge force in simulations can
either be overly smooth or grainy as compared to the actual self-force
in space charge dominated beams. Additional assumptions are employed
by different codes with respect to details of modeling the emission
process from the photocathode, whether or not the velocities of
individual electrons in the rest frame of the bunch are treated as
negligible, 3D or 2D or lower dimensionality nature (e.g. uniform
cylinders of HOMDYN) of the space charge, etc. The validity of these
assumptions must be evaluated for each individual case. Yet, data
comparing directly measured beam conditions, detailed phase space
distributions in particular, with simulations is sparse for space
charge dominated bunched beams such as found in either DC or RF
photoinjectors \cite{Cianchi, Miltchev, CHBB08}. Even a relatively
simple configuration involves a number of ``free'' parameters that
need to be varied within the uncertainty of the measurement such as
the RF phase in an RF gun in order to obtain good agreement between
simulations and measurements \cite{CHBB08}.

We present direct measurements of the transverse phase space
distribution using a simple beamline consisting of a DC gun followed
by an emittance compensation solenoid. The number of potential
variables affecting beam performance is reduced to the bare minimum in
such a setup. Careful characterization of the initial conditions such as
the laser transverse and temporal profiles and thermal emittance of the
photocathode allows us to carry out a direct cross-checking between the
measurements and 3D space charge simulations using the codes Parmela3D
\cite{Parmela} and GPT \cite{GPT}. Phase space distributions are then
used to calculate beam brightness vs.~the included beam fraction,
which is compared to the theoretical limit set by the thermal
transverse energy and accelerating field at the photocathode.

\begin{figure*}
\includegraphics[width=1.0\linewidth,clip]{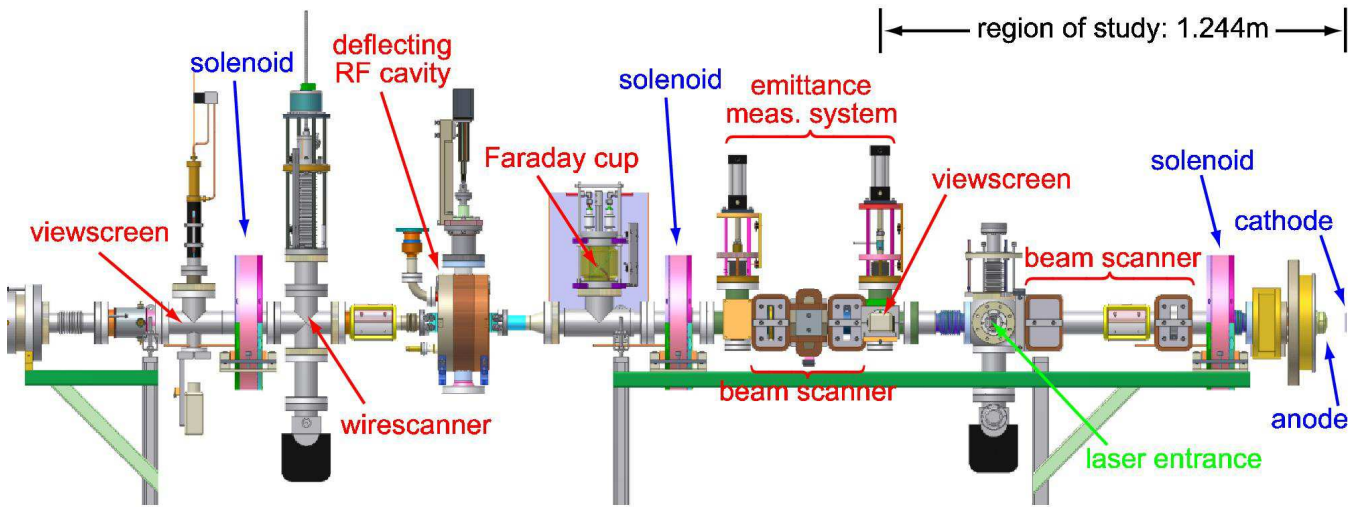}
\caption{\label{fig:beamline}Beamline used in the space charge
  studies. Beam direction is to the left.}
\end{figure*}

The paper is organized as following: Section~\ref{sec:setup} details the
experimental setup and beam diagnostics used in this work as well as our
experimental procedures. Section~\ref{sec:sims} presents simulation details and
data processing procedures used to extract information from the
measured data such as second moments and rms emittances. Comparison
between data and simulations follows in
Section~\ref{sec:results}. Finally, we conclude with a discussion and
outlook for future work.

\section{Experimental Setup\label{sec:setup}}
\subsection{Beamline}

\begin{figure}[b]
\includegraphics[width=1.0\linewidth,clip]{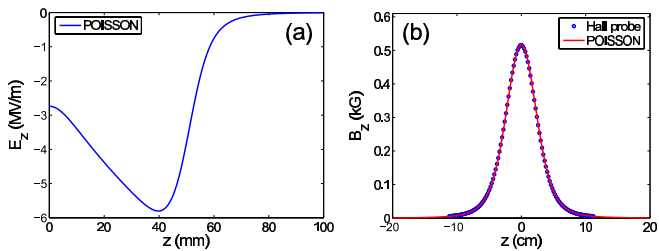}
\caption{\label{fig:fields}Field profiles for (a) the DC gun at 250 kV
  and (b) the solenoid with 5 A excitation current.}
\end{figure}

Fig.~\ref{fig:beamline} shows the experimental setup, which consists of
a high-voltage DC gun \cite{Dunham07} followed by a solenoid used for
emittance compensation located at 0.335 m from the photocathode to its
center. Fig.~\ref{fig:fields}b shows the distribution of the magnetic
field in the solenoid as calculated by POISSON \cite{POISSON} and the
actual measured values using a Hall-probe. Both the emittance
measurement system (EMS) and an insertable viewscreen are positioned
1.244 m from the photocathode. Additionally, the beamline is equipped
with a deflecting RF cavity \cite{Belomestnykh07} used to characterize
the initial temporal profile of photoemitted electrons, two beam
scanners and a Faraday cup all integrated into a data acquisition system for
direct phase space measurement. Details on the EMS are in the next
subsection. Two different types of materials have been employed for
viewscreens: high sensitivity BeO used in temporal measurements with
very low bunch charges and the deflecting cavity, and CVD diamond used
with average beam currents of up to 100 $\mu$A. Each viewport is
equipped with 12-bit CCD camera interfaced to the control system. 

The HV DC gun, initially designed for 750 kV, was operated at 250
kV. The field distribution is shown in Fig.~\ref{fig:fields}a. The gun
has reached 420 kV voltage during high voltage processing in the year
2007. Since then, however, we had to limit the gun voltage to a
conservative value below 300 kV due to field emission problems. Upon
disassembly of the gun we found a considerable amount of dust coming
from the ceramic resistive coating, which is believed to be the
primary reason behind the strong field emission. Work is underway to
eliminate this source of dust from the gun.

The laser system has been detailed elsewhere \cite{Ouzounov07,
Bazarov08b}. The laser spot-size was monitored using a 12-bit CCD
camera at the location of a virtual cathode.A Pockels cell was
used to reduce the 50 MHz train of pulses, with an average power of
about one Watt and 520 nm wavelength, to a lower duty factor for
beam measurements. A typical average current during phase space
measurements was between 10 to 100 $\mu$A. 

\begin{figure}
\includegraphics[width=1.0\linewidth,clip]{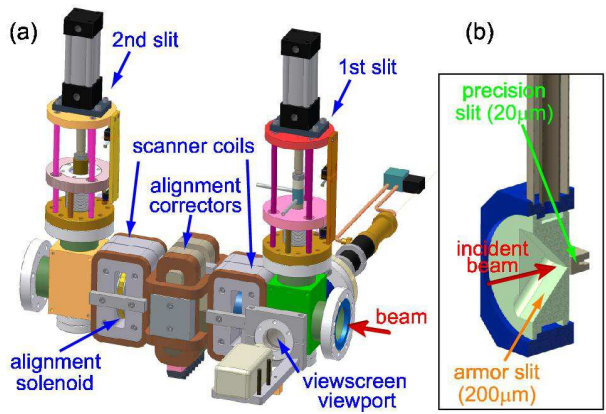}
\caption{\label{fig:ems}(a) Emittance Measurement System. One beam
  scanner (not shown) precedes the 1st slit. (b) Details of the 1st
  slit showing water-cooled 200 $\mu$m armor slit and 20 $\mu$m
  precision slit.}
\end{figure}

\subsection{Emittance Measurement System}
\begin{figure}
\includegraphics[width=1.0\linewidth,clip]{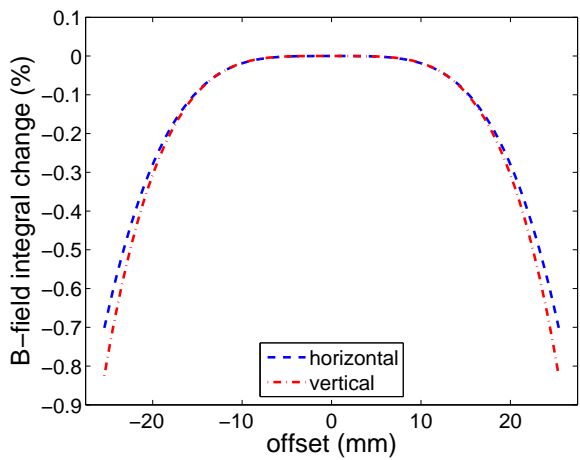}
\caption{\label{fig:scanner}Uniformity of B-field integral for beam
  scanner coils. Quantity being plotted is
  $\mathcal{I}(x,0)/\mathcal{I}(0,0)-1$ and
  $\mathcal{I}(0,y)/\mathcal{I}(0,0)-1$, where $\mathcal{I}(x,y) \equiv
  \int B_x(x,y,z)dz$ and $x = y = 0$ corresponds to the center of the
  beampipe.}
\end{figure}

Considerable care is required when designing a system for direct
measurement of the phase space of space charge dominated beam
\cite{Anderson02, Lejeune80}. The EMS employed in this study is a
double-slit system, with non-moving parts; see Fig.~\ref{fig:ems}. The
beam motion is achieved with a pair of corrector coils, designed with
a vanishing sextupole component in order to provide uniform kick
across most of the vacuum beampipe cross-section (see
Fig.~\ref{fig:scanner}). Each of the two scanners, one prior to the
1st slit, and the other between the two slits consists of a pair of
identical coils with opposite direction of excitation current. The
coils in each beam scanner have been measured to cancel each other to
better than 1\%. Thus, each scanner changes only the 
position of the beam without affecting its divergence. Additionally, a
pair of horizontal and vertical steering coils before each of the slits
allows correction for yaw/pitch alignment errors, leading to overall
relaxed tolerances for the device. Finally, a small solenoid (about
$\pm1^\circ$ rotation angle) is positioned between the two slits to
allow for roll compensation, although its use proved unnecessary in
practice.

The 1st EMS slit consists of an armor slit with a 200 $\mu$m opening
followed by a 20 $\mu$m vertically selecting precision slit brazed to
the water cooled armor slit; see Fig.~\ref{fig:ems}b. With this design most of
the beam power is intercepted by the armor slit. ANSYS analysis shows
that the EMS system is able to perform without significant deformation
of the precision slit ($<$ 10\%) with 1 kW of incident beam power.

Phase space measurements can be carried out using either single slit
and viewscreen or double-slit and Faraday cup configurations. Monte-Carlo
analysis on scattered radiation using GEANT4 \cite{GEANT4} has been
carried out for the full EMS system showing excellent signal to noise
performance of the system with either of the two configurations
\cite{Gulliford}.

All of the measurements reported in this work have been done using a
double-slit method. The maximum scan rate for the beam scanners was
200 Hz. The signal from both slits has been detected using the Faraday
cup connected to a low noise current amplifier. A solenoid positioned
just before the Faraday cup was used to focus the beamlet to the
center of the Faraday cup for increased charge collection
efficiency. A typical transverse phase space scan of 100$\times$100
points would take on the order of one minute.

\begin{figure}[b]
\includegraphics[width=1.0\linewidth,clip]{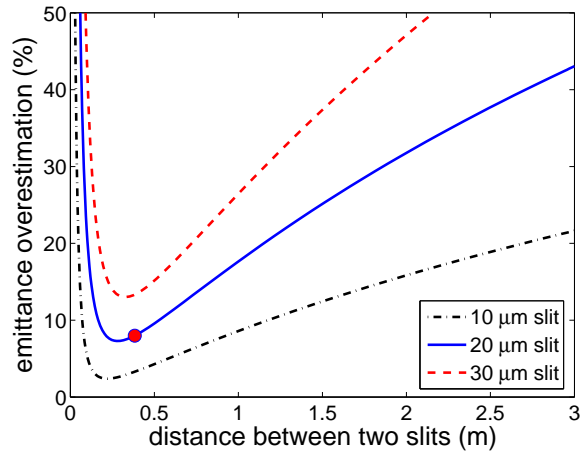}
\caption{\label{fig:sliterror}Emittance overestimation due to space
  charge for different slit openings. Beam parameters:
  $\epsilon_{n,x,y} = 0.3$ $\mu$m, rms bunch 
  duration $\sigma_t=20$ ps, $\sigma_{x,y} = 1.3$ mm, total rms
  divergence $\sigma_{T,x,y}' = 0.72$ mrad, kinetic energy 0.5 MeV,
  and bunch charge 80 pC. Red dot shows the actual separation between
  the two slits (38 cm).}
\end{figure}

The slit opening size and distance between the two slits have been
determined by solving coupled beam envelope equations for a beamlet
selected by the 1st slit with inclusion of the space charge force:
\begin{equation}\label{eq:bee}
\left\{
\begin{array}{ll}
\displaystyle \sigma_x''-\frac{I}{I_0(\beta\gamma)^3(\sigma_x+\sigma_y)} -
\frac{\epsilon_{n,x}^2}{\sigma_x^3(\beta\gamma)^2} & = 0,\\ 
\displaystyle \sigma_y''-\frac{I}{I_0(\beta\gamma)^3(\sigma_x+\sigma_y)} -
\frac{\epsilon_{n,y}^2}{\sigma_y^3(\beta\gamma)^2} & = 0. 
\end{array}
\right.
\end{equation}
Here $I$ is the beam peak current after passing through the 1st slit, $I_0$ =
17 kA is the Alfven current, and $(\beta\gamma)$ is the normalized
momentum. After a vertically selecting slit with opening $d$ small
compared to the beam size, one has $\sigma_y = d/\sqrt{12}$ and
normalized rms emittance $\epsilon_{n,y} =
\epsilon_{n,y0}(d/\sqrt(12))/\sigma_{y0}$, where $\epsilon_{n,y0}$ and
$\sigma_{y0}$ are the emittance and vertical size of the full beam
prior to the slit. By
solving Eq.~\ref{eq:bee} for the beamlet size $\sigma_y(L)$ at the
location $L$ of the 2nd slit (or the viewscreen), and comparing
$\sigma_y(L)/L$ to the uncorrelated divergence $\sigma_{y0}'$ at the
1st slit position, one can gauge the effectiveness of the slit system
in the presence of space charge. Here we note that the simple
requirement that the beamlet after the 1st slit be emittance
dominated \cite{Anderson02} is necessary but not sufficient. Because
the ratio of the space charge and emittance terms in Eq. (1) scales as
fast as $\propto \sigma_y^3$, meeting such condition at the location
of the 1st slit does not ensure that the beamlet stays emittance
dominated all the way through to the location of the 2nd
slit. Numerically solving the coupled equations Eq.~\ref{eq:bee}, on
the other hand, allows proper characterization of the slit
performance. Fig.~\ref{fig:sliterror} shows the results of emittance
overestimation by the double-slit method in the case of 80 pC bunches with
the parameters as indicated in the figure. The separation between
the two slits was chosen to be 38 cm leading to the overestimation in
emittance measurement to be less than 10\% of 0.3 $\mu$m normalized
rms emittance at 0.5 MeV kinetic energy.

\begin{figure}
\includegraphics[width=1.0\linewidth,clip]{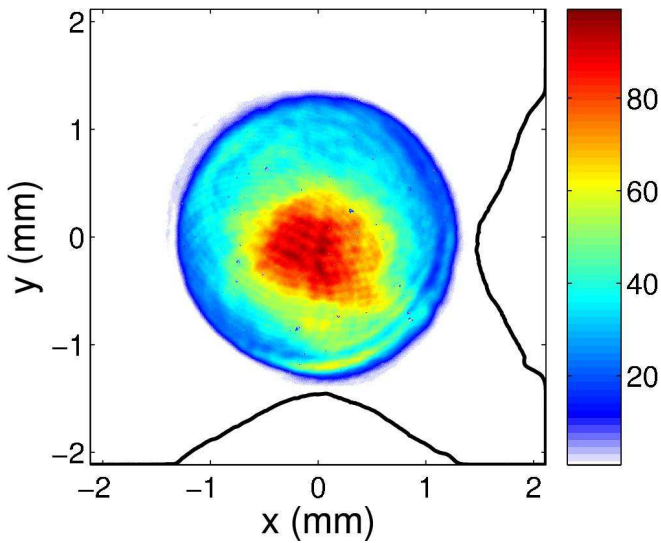}
\caption{\label{fig:laserxy}Typical transverse laser profile. Black
  solid lines show projected horizontal and vertical profiles.}
\end{figure}

\subsection{Experimental Procedures}
Measurements have been taken at 3 different bunch charges: 80 pC, 20
pC and 0.5 pC. The measured laser intensity stability was 2\% rms. The
laser spot was initially magnified, then passed through a 2.6 mm
diameter aperture, which was 1:1 imaged onto the photocathode. The
laser pointing stability was 60 $\mu$m rms in each transverse
direction. Each data set involved taking multiple images of the laser
spot on the virtual photocathode, and an image representing the average
centroid position was chosen for simulations as detailed in the next
section. A typical transverse laser spot profile is shown in
Fig.~\ref{fig:laserxy}.

\begin{figure}
\includegraphics[width=1.0\linewidth,clip]{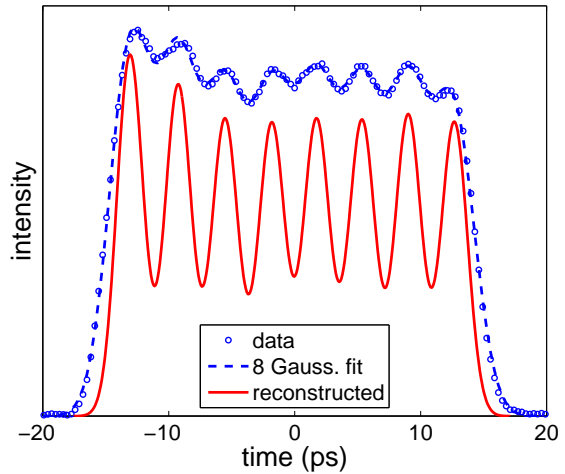}
\caption{\label{fig:lasert}Temporal profile of electron bunches as
  measured by the deflecting cavity with negligible charge per
  bunch. Red solid line shows the reconstructed profile used in
  simulations. Refer to the text for details.}
\end{figure}

The laser pulses were temporally stacked using three birefringent
crystals \cite{Bazarov08b}. Direct measurement of the initial temporal
distribution of electrons was performed with the deflecting cavity with
negligible charge per bunch. See Fig.~\ref{fig:lasert}. The resolution
of the temporal measurement in this case is 1.5 ps rms as limited by
the RF to laser synchronization and finite electron beam spot size. To
obtain the actual temporal profile of the electron distribution, the
data was fitted with 8 Gaussians, then each Gaussian was assigned 1.0
ps sigma corresponding to the value found in an autocorrelation
measurement for an unshaped laser pulse \cite{Ouzounov07}. Both the
fit to the data and reconstructed temporal profile used in simulations
are shown in Fig.~\ref{fig:lasert}.

Early on in the measurements, we were able to observe asymmetric
transverse phase space distribution, e.g.~see Fig.~\ref{fig:tail}. To
eliminate possible causes for such an occurrence, careful beam based
alignment was carried out before each data collection. A small laser
aperture (0.25 mm) placed concentrically with the larger one was used
to create a small beam with negligible charge per bunch. The beam
centroid vs.~the solenoid current data was fit to obtain both the angle and
the offset of the magnet's magnetic axis with respect to the
beam. The solenoid has been physically adjusted so that its axis
coincides with the beam axis to within a few 10's $\mu$m and $\mu$rad in
offset and angle respectively. Similarly, the center of the aperture
was imaged to coincide with the electrostatic center of the gun to
about 10 $\mu$m, ensuring that the central orbit is well aligned
throughout the system.

\begin{figure}
\includegraphics[width=1.0\linewidth,clip]{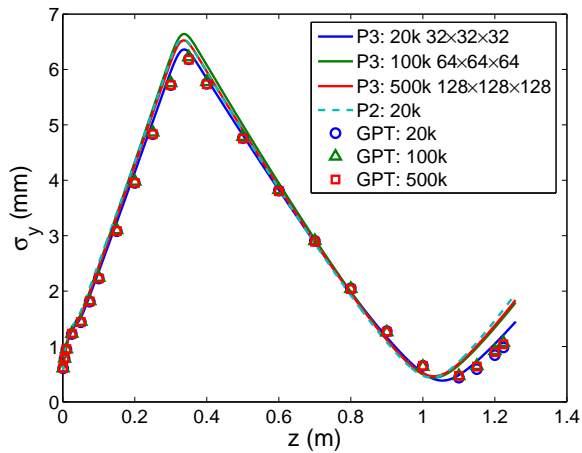}
\caption{\label{fig:sy_sim}Comparison of beam envelope
  vs. longitudinal position calculations for Parmela3D (solid lines)
  and GPT (open symbols) for 80 pC charge per bunch. Dashed line shows
  Parmela calculations using 2D space charge routine. The solenoid
  current is 3.6A. The gun voltage is 250 kV.}
\end{figure}

\begin{figure}
\includegraphics[width=1.0\linewidth,clip]{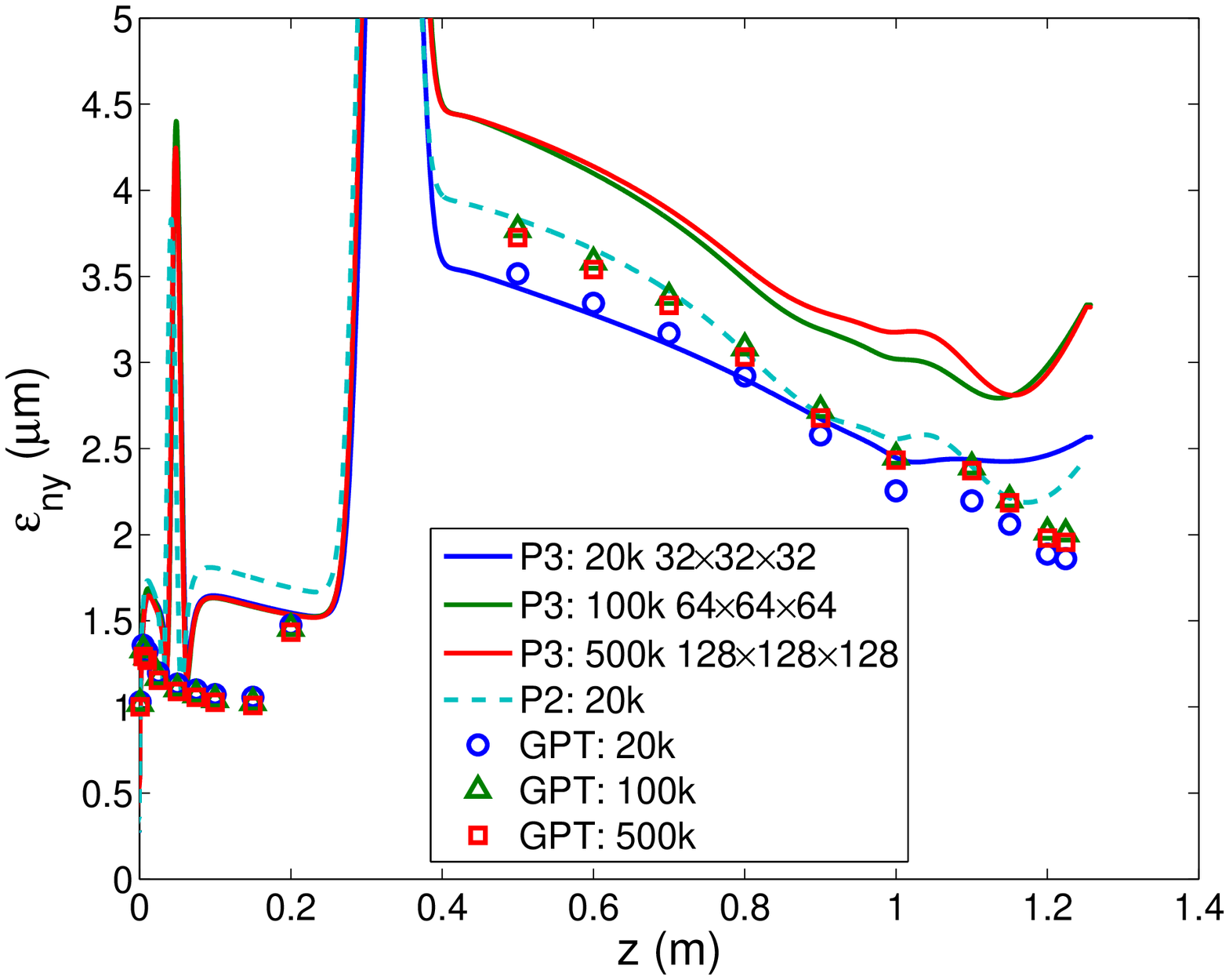}
\caption{\label{fig:eny_sim}Comparison of emittance vs.~longitudinal
  position calculations for Parmela3D (solid lines) and GPT (open
  symbols) for 80 pC charge per bunch. Dashed line shows Parmela
  calculations using 2D space charge routine. The solenoid current is
  3.6A. The gun voltage is 250 kV.}
\end{figure}

\begin{figure}
\includegraphics[width=1.0\linewidth,clip]{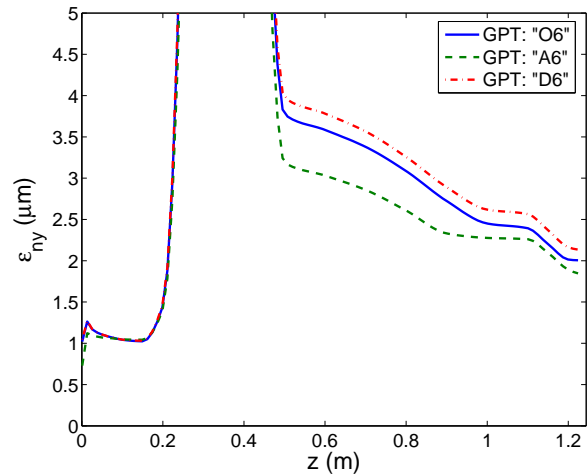}
\caption{\label{fig:bc} Comparison of emittance vs. position
  calculations for GPT in case of different boundary conditions: open
  (O6), approximate (A6), and Dirichlet (D6). Refer to the text for
  details. Bunch charge is 80 pC, the solenoid current is 3.6A. The
  gun voltage is 250 kV. 100k macroparticles were used in the
  simulations.}
\end{figure}

\section{Simulation Parameters And Data Processing
  Procedures\label{sec:sims}} 
\subsection{Simulation Parameters And Conditions}
As the quantum efficiency (QE) of the photocathode in the region of 
interest was found to have about 10\% peak-to-peak fluctuations (in
absolute terms the QE for GaAs was about 6\% in this work), the
measured laser transverse profiles and the temporal shape
shown in Fig.~\ref{fig:lasert} were used to created 3D distributions for
simulations. The thermal emittance of GaAs photocathodes has been
investigated previously \cite{Bazarov08a} and for an rms laser spot
size 
$\sigma_\perp$ was found to be $\epsilon_{n,th} =
\sigma_\perp \sqrt{kT_\perp/mc^2}$ with $kT_\perp = 120\pm 8$ meV for
520 nm light, $mc^2$ is electron rest energy.

We have used two 3D space charge codes: Parmela3D which
employs a fast Fourier Transform method for solving the Poisson
equation on a 3D grid \cite{Young03} and General Particle Tracer (GPT)
with a non-equidistant mesh solver for the space charge force
calculation \cite{vanderGeer:2003zb}. The same field maps for both the
gun and 
the solenoid magnet were used in both codes. Convergence of the
calculation results has been checked for 20k, 100k, and 500k
particle distributions and different mesh sizes. Results for
the beam envelope are presented in
Fig.~\ref{fig:sy_sim}. Fig.~\ref{fig:eny_sim} shows the results for
emittance vs.~longitudinal position for the case of 80 pC bunches. An
additional difference between Parmela and GPT is that the former
reports the relevant beam parameters as a function of time, whereas 3D
coordinates of the bunch were projected to a given longitudinal
position for the latter. We observe that sufficient 
convergence is demonstrated with 100k macroparticles for both
Parmela3D and GPT. The mesh size was set to $64\times64\times64$ for
Parmela3D and $50\times50\times50$ for the non-equidistant mesh
Poisson equation solver in GPT with the bounding 
box size set to $5\sigma$ in each dimension. Additionally, the Poisson
equation solver in GPT provides a choice of 3 different boundary
conditions at the bounding box: Dirichlet with zero potential, an open
boundary, and an approximate boundary in which the potential at the
bounding box is assigned analytically computed values from a uniform
elliptical cylinder with rms dimensions set equal to those of the actual
bunch. (Zero potential is assigned at the cathode to include image
charge effects for all boundary condition types). Comparison between
different boundary conditions is shown in Fig.~\ref{fig:bc} for emittance
vs.~longitudinal position. Beam envelope calculations for these
boundary conditions produce essentially identical results for these
parameters, and therefore are not shown. Open boundary
conditions were used for subsequent comparisons with the measurements.

\begin{figure}
\includegraphics[width=1.0\linewidth,clip]{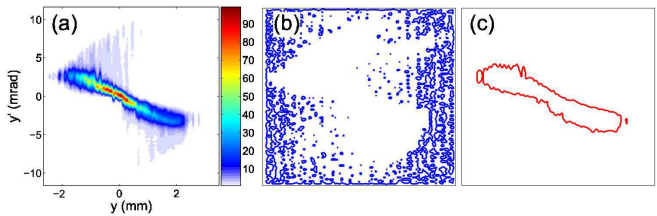}
\caption{\label{fig:noisethresh}Example of measured transverse phase
  space (a), with a contour map of a binary image after applying 0.6\%
  (b) and 4.4\% (c) threshold of the peak intensity.}
\end{figure}

\begin{figure}
\includegraphics[width=1.0\linewidth,clip]{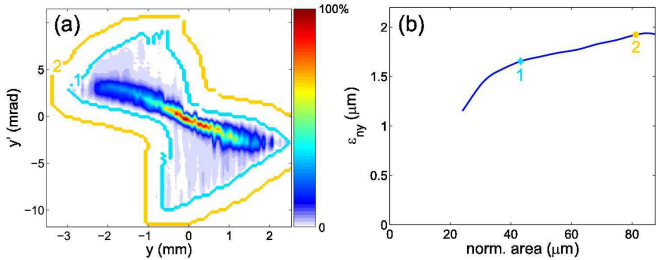}
\caption{\label{fig:noisesub}Noise subtraction verification
  procedure. (a) Example transverse phase space with the contour (1)
  obtained through the boundary detection algorithm and the grown
  contour (2), which corresponds to 50\% of the 
  available data treated as noise. (b) Normalized rms emittance
  calculation as a function of included area after the noise
  subtraction. Refer to the text for details.}
\end{figure}

\subsection{Data Processing Procedures}
In order to extract second moments from the measured beam
profiles and phase space distributions, the data requires
appropriate noise subtraction. The general approach follows
the general notion of self-consistent unbiased rms emittance
analysis (SCUBEEx) \cite{Stockli04}: i) a certain contour delineates
the signal plus noise region from the noise only region; ii) the
average intensity of the outside region represents a noise bias; iii)
the noise bias is subtracted from the data while the outside region is
assigned 0 intensity; iv) the contour is grown, and the parameter of
interest should not change significantly as a function of the included
area once all signal is accounted for if uniform random noise is
present. We have used two types of contours: circular type for the
viewscreen data, and a special boundary detection technique for the
measured phase-space distributions. The boundary detection technique
is based on the following observation: a binary image obtained by
applying a threshold to the phase space 2D distribution is likely to
form a continuous region for the signal, and many individual islands
for the noise. See Fig.~\ref{fig:noisethresh}. The boundary detection
algorithm proceeds as follows: i) the data is convolved with a
$n\times n$ square (image blurring); ii) the smallest threshold is
found that generates a single continuous island; iii) $n$ is
incremented and step i) is repeated. The process stops when the island
starts to include chunks of noise region, which becomes clearly
visible. Once the boundary has been found, noise subtraction is
verified by growing/shrinking the contour. The contour growing is
stopped when less than half of the whole image area becomes available
for noise estimation. Fig.~\ref{fig:noisesub} illustrates the
procedure further. The change in the parameter of interested (e.g. rms
emittance) vs. the included area delineating the signal from noise
regions represents the uncertainty in the measurement due to the noise
subtraction. Emittance or rms values so calculated correspond to 100\%
of the beam.

\begin{figure}
\includegraphics[width=1.0\linewidth,clip]{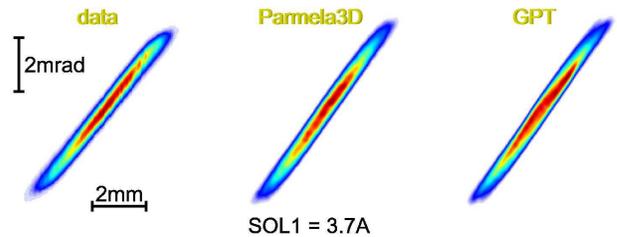}
\caption{\label{fig:05pC}Comparison of measured and simulated vertical
  transverse phase space distributions for 0.5 pC bunches at z = 1.244
  m. Solenoid current is 3.7 A. Corresponding rms normalized
  emittances: $\epsilon_{n,y} = 0.31 \pm 0.04$ $\mu$m (data), 0.29
  $\mu$m (Parmela3D), 0.28 $\mu$m (GPT).  Corresponding rms sizes:
  $\sigma_y = 1.15 \pm 0.05$ mm (data), 1.14 mm (Parmela3D), 1.14 (GPT).}
\end{figure}

\begin{figure*}
\includegraphics[width=1.0\linewidth,clip]{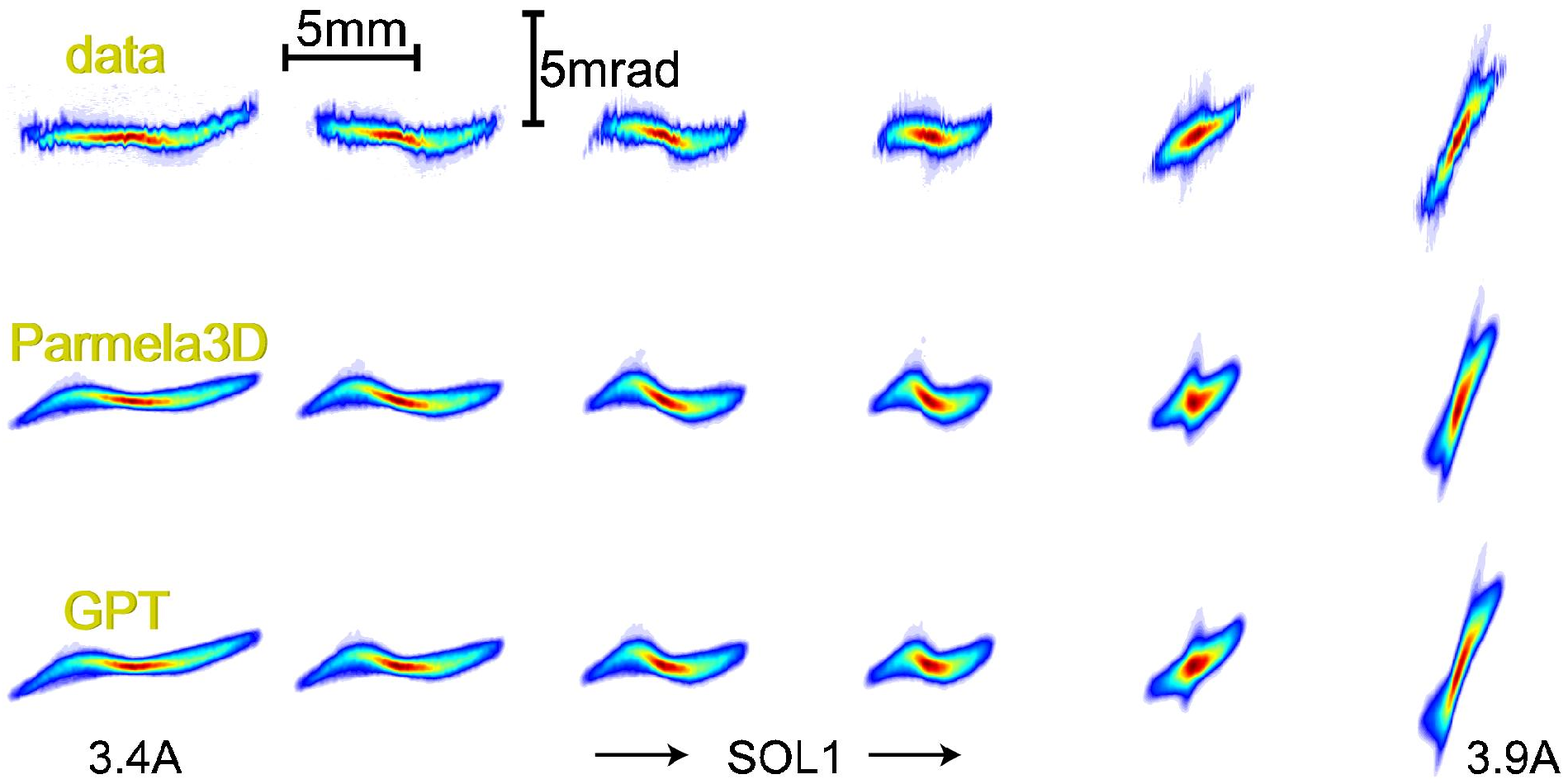}
\caption{\label{fig:20pC}Comparison of measured and simulated vertical
  transverse phase space distributions for 20 pC bunches at z = 1.244
  m. Data representing measurements, Parmela3D and GPT calculations is
  arranged in rows with different strength of the solenoid lens
  corresponding to column position.}
\end{figure*}

\begin{figure*}
\includegraphics[width=1.0\linewidth,clip]{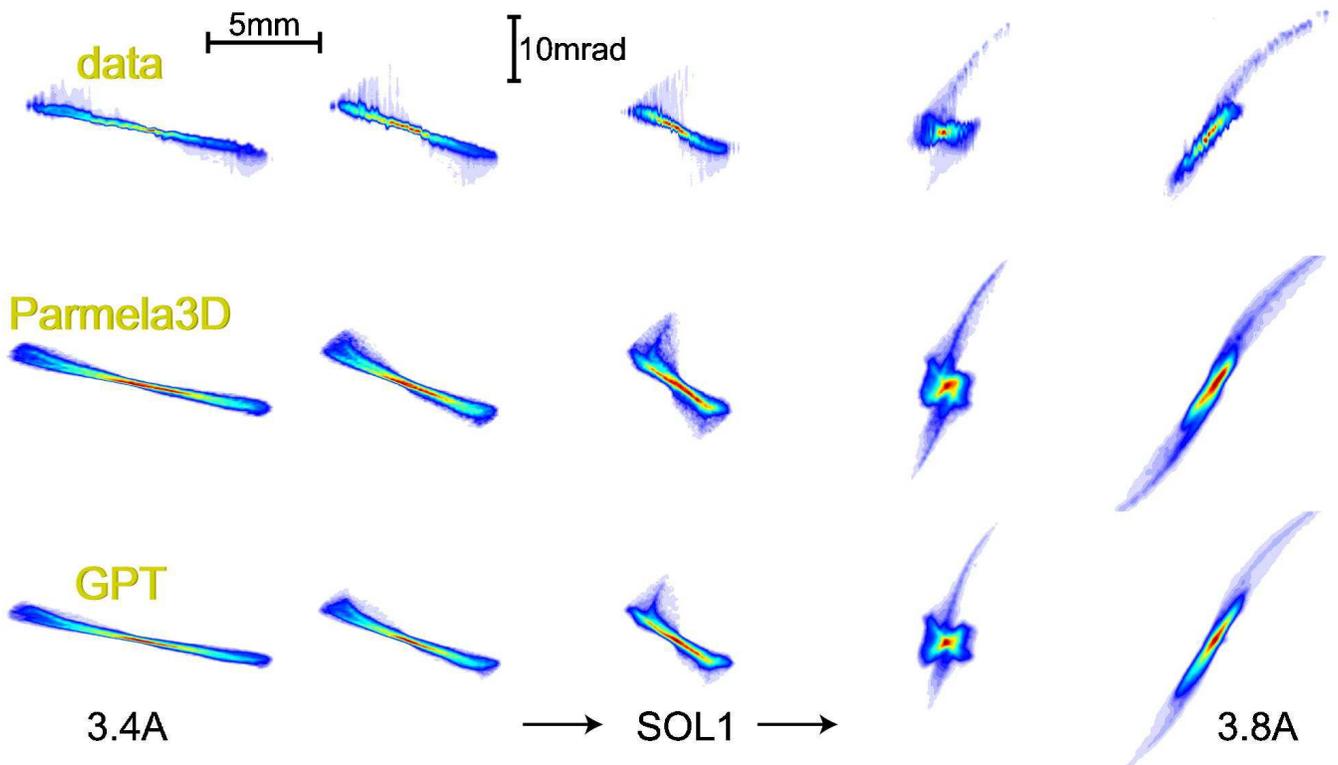}
\caption{\label{fig:80pC}Comparison of measured and simulated vertical
  transverse phase space distributions for 80 pC bunches at z = 1.244
  m. Data representing measurements, Parmela3D and GPT calculations is
  arranged in rows with different strength of the solenoid lens
  corresponding to column position.}
\end{figure*}

\section{Comparison Of Measurements With
  Simulations\label{sec:results}}
\subsection{Transverse Phase Space Distributions}
Fig.~\ref{fig:05pC} shows the comparison of measured transverse phase
space at the location of the 1st slit (z = 1.244 m) for 0.5 pC charger
per bunch with Parmela3D and GPT simulations. The resolution of the
measured transverse phase space is $90\times90$ steps. In the case of
simulations, each image is produced using a $300\times300$ 2D
histogram with additional convolution (blurring) with a $3\times3$
square. Each image is normalized to the same maximum intensity
value. An identical color map to that of Fig.~\ref{fig:laserxy} is
used throughout. As expected, the calculated rms normalized emittance
is in good agreement with the thermal emittance value for this case.

\begin{figure}
\includegraphics[width=1.0\linewidth,clip]{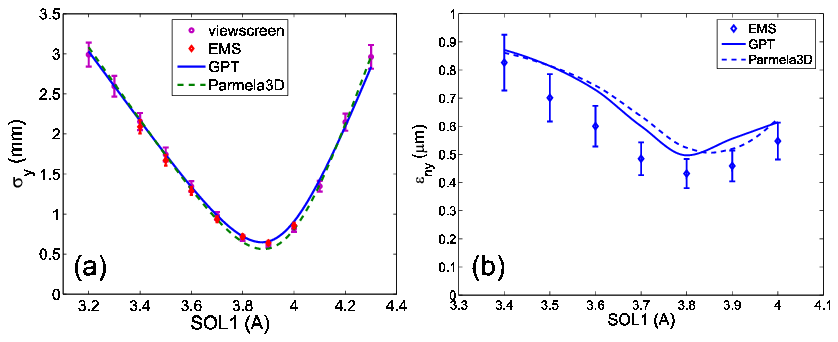}
\caption{\label{fig:20pCsm}Comparison of vertical rms beam size (a) and
  normalized rms emittance (b) at the location of the measurement z =
  1.244 m vs. solenoid excitation current for 20 pC bunches.} 
\end{figure}

\begin{figure}
\includegraphics[width=1.0\linewidth,clip]{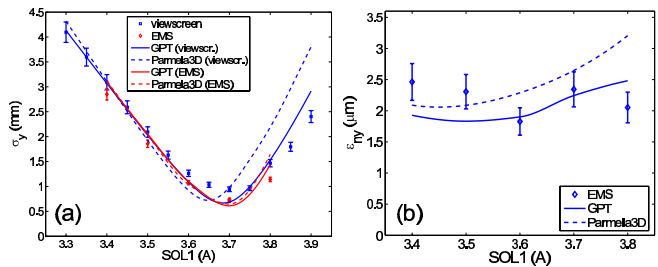}
\caption{\label{fig:80pCsm}Comparison of vertical rms beam size (a)
  and normalized rms emittance (b) at the location of the measurement
  z = 1.244 m vs. solenoid excitation current for 80 pC bunches.} 
\end{figure}

Fig.~\ref{fig:20pC} and \ref{fig:80pC} show the comparison for 20 pC
and 80 pC charge per bunch respectively vs.~solenoid lens
strength. The streak features seen in the measured phase space are due
to the motion of the laser spot at the photocathode. Good qualitative
agreement can be seen for 20 pC/bunch data, while some discrepancy in
the shape of transverse phase space distributions can be seen at
larger solenoid current values for 80 pC/bunch.

\subsection{Second Moments Of The Beam}
Fig.~\ref{fig:20pCsm} shows a comparison of vertical rms beam size (a) at the
location of the viewscreen (z = 1.244 m) and rms normalized
vertical emittance for 100\% of the beam (b) as a function of
solenoid current for 20 pC bunches. Excellent agreement is seen for
the spot size comparison and good overall agreement for the emittance
values, although the measured rms emittance appears to be
systematically smaller for the 20 pC/bunch case.

Fig.~\ref{fig:80pCsm} shows similar results for 80 pC charge bunches.
Different sets of curves for EMS and viewscreen simulations correspond
to different laser spots as registered for the two data sets. Good
agreement between simulations and measurements of the beam size is
seen for 80 pC/bunch before the formation of a beam waist at the
location of measurement, while the agreement at larger solenoid
currents appears to be less conclusive.

\begin{figure}
\includegraphics[width=1.0\linewidth,clip]{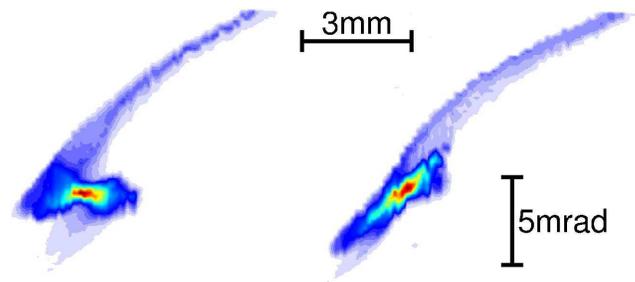}
\caption{\label{fig:tail}Example of asymmetric phase space
  distribution measured for 80 pC/bunch at the solenoid current of
  3.7 A (left) and 3.8 A (right). The top part of the asymmetric tail
  was truncated during the measurement.}
\end{figure}

\begin{figure}
\includegraphics[width=1.0\linewidth,clip]{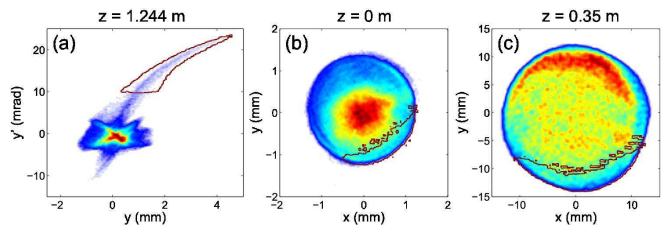}
\caption{\label{fig:tailsim}Calculated vertical transverse phase space
  using GPT (a) at the location of the 1st slit (z = 1.244 m) for a
  given transverse initial laser spot size (b). Transverse profile is
  shown at the location inside the solenoid (c), z = 0.35 m. Note two
  different scales for (b) and (c). The portion of the distribution
  responsible for the tail is also shown.}
\end{figure}

\subsection{Asymmetric Phase Space Distributions}
Asymmetric phase space distributions have been measured on multiple
occasions for space charge dominated beam conditions in our setup. For
example, see Fig.~\ref{fig:tail}. No such asymmetry was observed for low bunch
charge running under otherwise identical operation conditions. It was
important to understand the origin of this behavior for its subsequent
mitigation. We were able to reproduce similar phase space
distributions in 3D simulations for laser spots with noticeable
asymmetry. To elucidate the mechanism for this tail formation, the
particles comprising the tail have been tagged and their portion of
the distribution is shown in Fig.~\ref{fig:tailsim} for the transverse
profile at the location of the photocathode, z = 0 (b), and inside the
solenoid, z = 0.35 m (c). The asymmetry in the laser spot causes the
space charge forces to push these particles away from the central axis
(three is about 2 mm difference between the top and
the bottom edges of the transverse distribution in
Fig.~\ref{fig:tailsim}c) so that the particles experience a stronger
focusing kick from the solenoid lens. These particles then undergo a
cross-over and form the observed 
phase space tail. Improving the transverse laser shape would reduce
the asymmetry in the phase space.

\begin{figure}
\includegraphics[width=1.0\linewidth,clip]{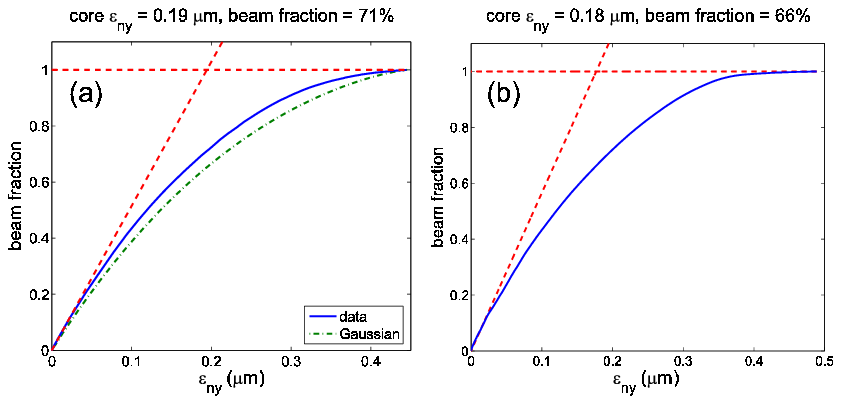}
\caption{\label{fig:20pCemitfrac}Normalized rms emittance vs. included
  beam fraction for measured (a) and calculated by GPT (b) phase space
  distributions for 20 pC/bunch. A corresponding Gaussian beam is also
  shown for comparison. The 100\% normalized rms emittance is
  $0.43\pm0.05$ $\mu$m for the measurement (a) and 0.49 $\mu$m for GPT
  (b).}
\end{figure}

\begin{figure}
\includegraphics[width=1.0\linewidth,clip]{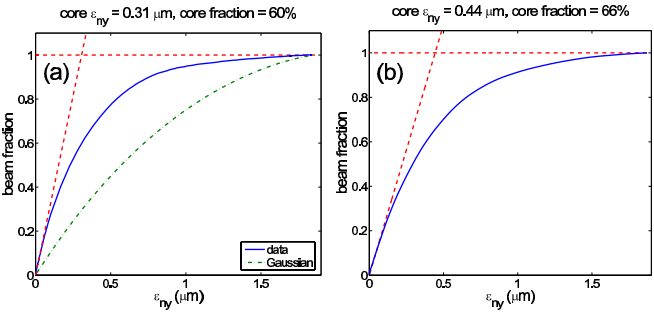}
\caption{\label{fig:80pCemitfrac}Normalized rms emittance vs. included
  beam fraction for measured (a) and calculated by GPT (b) phase space
  distributions for 80 pC/bunch. A corresponding Gaussian beam is also
  shown for comparison. The 100\% normalized rms emittance is $1.8 \pm
  0.2$ $\mu$m for the measurement (a) and 1.8 $\mu$m for GPT (b).}
\end{figure}

\subsection{Theoretical Limit To Beam Brightness}
It is instructive to consider rms emittance as a function of the
contributing beam fraction \cite{Lejeune80}. For example,
Fig.~\ref{fig:20pCemitfrac} shows rms normalized emittance vs.~beam
fraction as measured and simulated using GPT for 20 pC/bunch for a
solenoid current of 3.8 A. Fig.~\ref{fig:80pCemitfrac} shows
corresponding results for 80 pC/bunch for a solenoid current of 3.6
A. Dot-dashed lines show the expected curve for a Gaussian
distribution in the 
phase space with the same rms emittance as 100\% of the actual
beam. It can be seen that in the case of 80 pC/bunch, the beam
strongly deviates from Gaussian distribution, having a substantially
brighter core. Core emittance, defined as $\epsilon_{n,y,core} \equiv
d\epsilon_{n,y}(\xi = 0)/d\xi$ with $\epsilon_{n,y}(\xi)$ being the
normalized rms emittance as a function of beam fraction $0 \leq \xi
\leq 1$, is given in both figures along with the core fraction
$\xi_{core}$ defined as the fraction of the beam with the emittance
equal to the core emittance value: $\epsilon_{n,y}(\xi_{core}) =
\epsilon_{n,y,core}$.

\begin{figure}
\includegraphics[width=1.0\linewidth,clip]{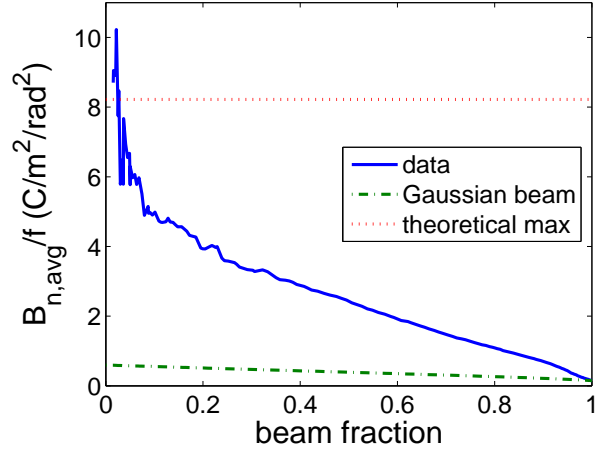}
\caption{\label{fig:br}Beam brightness normalized per single bunch as
  a function of contributing beam fraction for 80 pC charge per
  bunch. Corresponding result for an equivalent Gaussian beam with the
  same 100\% rms emittance is shown for comparison (dot-dashed
  line). Dotted line shows theoretical limit as given by 
  Eq.~\ref{eq:maxbr}. The rms normalized emittance is $1.8 \pm 0.2$
  $\mu$m.}
\end{figure}

The beam brightness available from photoinjectors forms through an
interplay of several phenomena such as space charge dominated beam
dynamics in the presence of time transient and position dependent
external fields. The upper limit, however, is set by the thermal
emittance of the photocathode and the available accelerating
gradient. Consider a short laser pulse illuminating a photocathode
placed in the accelerating field $E_{cath}$ . The electron bunch after
the emission will assume a pan-cake shape provided that the laser
pulse duration is sufficiently short: $\sigma_t \ll \sqrt{\sigma_\perp
m/e E_{cath}}$, $m$ and $e$ are electron mass and charge
respectively. This condition is satisfied in most operating
photoinjectors with bunched beams. The maximum charge density that can
be supported by the electric field is then given by
\begin{equation}\label{eq:sclimit}
\frac{dq}{dA} = \epsilon_0 E_{cath}.
\end{equation}
Note the inclusion of the image charge. The average (normalized) beam
brightness can be defined as a ratio of average current $I_{avg}$ over
its 4D-volume $\mathcal{A}_4$ defined for $(x,p_x/mc,y,p_y/mc)$
coordinates:
\begin{equation}\label{eq:brdef}
\mathcal{B}_{n,avg} = \frac{I_{avg}}{\mathcal{A}_4}.
\end{equation}
This quantity would be related to x-ray brightness for a properly
matched undulator, for example. Beam brightness normalized per single
bunch is given by $\mathcal{B}_{n,avg}/f = q/\mathcal{A}_4$, with $f$
being the repetition rate, and $q$ the charge contained in the
4D-volume $\mathcal{A}_4$. E.g.~the 4D volume $\mathcal{A}_4 =
dxdp_xdydp_y/(mc)^2$ for a 4D-hypercuboid element with sides $dx$,
$dp_x$, $dy$, and $dp_y$. The charge contained in $\mathcal{A}_4$ can
be written as
\begin{equation}\label{eq:charge}
\frac{dq}{dA}dxdp_xdydp_y\frac{1}{\kappa\sigma_p^2},
\end{equation}
with $\sigma_p$ being the rms value of the transverse momentum
(assumed to be isotropic for both transverse directions), which is
$\sigma_p = \sqrt{mkT_\perp}$ for Maxwell-Boltzmann distribution of
velocities. The dimensionless coefficient $\kappa$ depends on details
of momentum distribution, e.g. $\kappa = 4\pi$ corresponds to a
uniform circular distribution in $p_x$ and $p_y$ with a diameter
$4\sigma_p$, while $\kappa = 2\pi$ corresponds to the peak of a 2D
Gaussian distribution. Combining Eq.~\ref{eq:charge} with
the charge density as given by Eq.~\ref{eq:sclimit}, we find the maximum beam
brightness normalized per single bunch 
\begin{equation}\label{eq:maxbr}
\frac{\mathcal{B}_{n,avg}}{f} = \frac{\epsilon_0 mc^2}{\kappa}\frac{E_{cath}}{kT_\perp}.
\end{equation}

This result shows the maximum beam brightness available from a
photoinjector to be independent from the bunch charge, and is
determined by the accelerating field $E_{cath}$ and transverse thermal
energy $kT_\perp$ of the electrons leaving the photocathode. To
compare Eq.~\ref{eq:maxbr} with the measured data, we compute the beam
brightness per single bunch as
\begin{equation}\label{eq:brmeas}
  \left.\frac{\mathcal{B}_{n,avg}}{f}\right|_{meas.} = q\left(\frac{\xi}{4\pi\epsilon_{n,y}(\xi)}\right)^2,
\end{equation}
where  is 100\% charge, and $\epsilon_{n,y}(\xi)$ is emittance
vs.~beam fraction curve
(cf.~Fig.~\ref{fig:80pCemitfrac}. Eq.~ref{eq:brmeas} additionally
assumes an axially symmetric beam with uniform phase space
distribution inside an equivalent ellipse with
$4\pi\epsilon_{n,y}(\xi)$ area. Fig.~\ref{fig:br} shows brightness
normalized per single bunch vs.~the beam fraction. The theoretical
(average) brightness maximum as given by Eq.~\ref{eq:maxbr} is shown
as well ($\kappa = 4\pi$). Additionally, $\mathcal(B)_{n,avg}/f$ is
computed for a beam that has the same 100\% rms emittance as the
actual beam but adopts a 2D Gaussian distribution in the phase
space. It is seen that an equivalent Gaussian beam does a poor job of
describing the 80 pC/bunch beam as the core is substantially brighter for
the measured beam approaching the limit given by Eq.~\ref{eq:maxbr}.

\begin{figure}
\includegraphics[width=1.0\linewidth,clip]{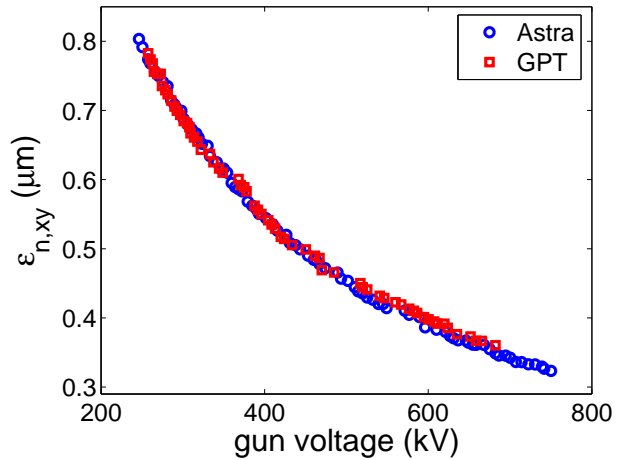}
\caption{\label{fig:optimum}Calculated normalized rms emittance (100\% of
  the beam) at z = 1.244 m vs.~the gun voltage for the beamline setup
  used in this study. Refer to the text for details on simulation parameters.}
\end{figure}

\section{Discussion And Outlook}
Benchmarking of 3D space charge codes has been performed with the
direct measurements of the transverse phase space for a bunched beam
in the space-charge dominated regime from a DC gun. Overall, good
agreement has been found between the measurements and simulations. We
observe that an equivalent Gaussian beam assigned the measured 100\%
rms emittance poorly describes the peak brightness available in the beam
at 80 pC bunch charge due to the presence of a substantially brighter
core. In particular, for the case of the minimum measured 100\% rms
normalized emittance  $\epsilon_{n,y} = 1.8\pm0.2$ $\mu$m, the core
emittance is found to be $\epsilon_{n,y,core} = 0.31\pm0.04$ $\mu$m
with 60\% beam fraction contained in the core. Additionally, the
comparison of the measured beam brightness vs.~beam fraction shows
that it approaches the maximum theoretical brightness as set by the
available accelerating gradient and transverse thermal energy of the
photocathode. While the peak brightness of the beam cannot be improved
without changing $kT_\perp$ and $E_{cath}$, it should be possible to
bring a larger portion of the electron beam to approach the brightness
limit given by Eq.~\ref{eq:maxbr} through proper control of the
space charge forces. For example, Fig.~\ref{fig:optimum} shows
simulated possible rms normalized emittance (100\% of the beam) from
the same beamline as used in this experiment for 80 pC bunches as a
function of gun voltage. A
uniform cylindrical laser distribution has been used in these
calculations, with the (same) thermal transverse energy corresponding
to GaAs illuminated by 520 nm. The laser pulse duration in these
simulations was 12 ps rms, about 50\% longer that what was used in the
measurements reported here. In addition to employing a longer laser
pulse and continuing the work to reach the gun design voltage of 750
kV, noticeable improvements are sought for the transverse laser shape
as well as the pointing stability. E.g. based on simulations, the
worst 10\% emittance beam fraction in our measurements can be mostly
attributed to the less than ideal transverse laser profile (see
Fig.~\ref{fig:laserxy}).

Finally, we note that despite a significantly more complicated setup
and beam dynamics in the full ERL injector \cite{Bazarov05}, where the
bunch undergoes acceleration to over 10 MeV, bunch compression and
matching into the linac, a much simpler beamline such as the one
considered in this study allows exploration of the best beam
brightness achievable from the whole photoinjector. In particular,
simulations for the full injector \cite{Bazarov07}, where
substantially shorter bunches are produced though subsequent drift
bunching ($\sim3$ ps rms) indicate rms normalized emittances at
$\sim11$ MeV which are about 50\% lower but otherwise very similar to
that shown in Fig.~\ref{fig:optimum}. Thus, to continue the work on
improving the HV DC gun design and pushing for lower emittances, it is
sufficient in many ways to have a simpler setup with beam diagnostics
dedicated to such research similar to the one described in this work.

\section{Acknowledgments}
We acknowledge control system support by John Dobbins, John Barley,
and Mike Forster, laser support by Dimitre Ouzounov and Heng Li,
magnetic measurements of the solenoid by Alexander Temhykh. The
following individuals are acknowledged for many useful discussions and
their interest in this work: Tsukasa Miyajima, Georg Hoffstaetter, and
Dave Rice. This work is supported by the NSF grant PHY-0131508 and
NSF/NIH-NIGMS award DMR-0225180.

\bibliography{paper}

\end{document}